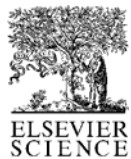
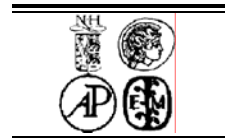

# Resonance Anomalous Surface X-ray Scattering

Andreas Menzel, Kee-Chul Chang, Vladimir Komanicky, Hoydoo You[*]

*Argonne National Laboratory, Materials Science Division, 9700 S. Cass Ave. Argonne, IL 60439, USA*

Yong S. Chu

*Argonne National Laboratory, Experimental Facility Division, 9700 S. Cass Ave. Argonne, IL 60439, USA*

Yuriy V. Tolmachev

*Kent State University, Chemistry Department, P. O. Box 5190, Kent, OH 44242, USA*

John J. Rehr

*University of Washington, Department of Physics, P.O. Box 351560, Seattle, WA 98195, USA*

**Abstract**

Resonance anomalous surface x-ray scattering (RASXS) technique was applied to electrochemical interface studies. It was used to determine the chemical states of electrochemically formed anodic oxide monolayers on platinum surface. It is shown that RASXS exhibits strong polarization dependence when the surface is significantly modified. The polarization dependence is demonstrated for three examples; anodic oxide formation, sulfate adsorption, and CO adsorption on platinum surfaces. $\sigma$- and $\pi$- polarization RASXS data were simulated with the latest version of *ab initio* multiple scattering calculations (FEFF8.2). Elementary theoretical considerations are also presented for the origin of the polarization dependence in RASXS.
© 2001 Elsevier Science. All rights reserved

*Keywords:* Surface x-ray scattering; Anomalous x-ray scattering; electrochemical interfaces; Dichroism
*PACS:* Type your PACS codes here, separated by semicolons ;

## 1. Introduction

Surface x-ray scattering has become a standard x-ray technique of studying the structures of surfaces and interfaces. Robinson and Tweet (1992) gave a general introduction to the technique and reviewded some details of the measurement procedures. Briefly, the x-rays diffracting from a surface or interface experience a sharply changing electron density at the surface or interface. As a result, in addition to the ordinary three-dimensional reciprocal lattice, x-rays diffract to rods of diffuse scattering, emanating from the origin and the other reciprocal lattice points of the reciprocal lattice. The remarkable sensitivity of the synchrotron x-rays to mono- and submono-layer absorbates as well as to reconstructions of surfaces and interfaces made the technique become one of the most popular x-ray tools in the past decades. In particular, the penetrating power of x-rays makes the technique uniquely applicable for buried interface studies where no electron-based surface techniques could be used.

Numerous variations of surface x-ray scattering were developed over the years depending on which part of the reciprocal space is measured or whether the surface is of single crystals, multilayers, liquid, or polymers. However, all of them can be categorized into three types of surface scattering, namely, x-ray reflectivity (specular and off-specular), glancing angle in-plane x-ray diffraction, and crystal truncation rod (CTR) measurements. Their experimental geometries are schematically shown in Fig. 1. The specular and off-specular x-ray reflectivity techniques (a) are used to study surface morphology. They are not typically sensitive to crystalline structures of the surface but sensitive to nano- to micro-meter length scales. Glancing angle in-plane diffraction (b), measured parallel to the surface with incident and diffracted x-rays at glancing angles, is extremely sensitive to the two-dimensional structure and reconstruction. Crystal truncation rods (c), measured normal to the surface, are extremely sensitive to the surface coverage and relaxation.

---

[*] Corresponding author. Tel.: 630-252-3429; fax: 630-252-7777; e-mail: hyou@anl.gov.

On the other hand, resonance scattering (more commonly referred to as anomalous scattering) has been used in crystallography over decades for its chemical or elemental sensitivity (e.g., Fanchon and Hendrickson, 1991) and in diffraction anomalous fine structure (DAFS) study (Stragier, 1992) for local structures as well as the chemical sensitivity. Naturally, there have been several studies incorporating the anomalous scattering or resonance phenomena of variable-energy synchrotron x-rays to the surface scattering techniques. Early studies for elemental sensitivity of the surface composition include the study of semiconductor surfaces (Walker et al., 1991) and interfaces (Akimoto et al., 1991) and electrode surfaces under water (Tidswell et al., 1995). Specht and Walker (1993) made first chemically sensitive measurement of the interfaces of an oxide heterostructure, $Al_2O_3/Cr_2O_3$, where they demonstrated the anomalous fine structure of CTR analogous to DAFS of bulk reciprocal lattices (Stragier, 1992). Later, Chu et al. (1999), You et al. (2000), and Park et al. (2005) have shown that the surface resonance x-ray scattering is indeed sensitive to the surface chemical state and can be modeled by combining the usual surface diffraction equation and the resonance anomalous real and imaginary scattering factors. The surface-sensitive anomalous scattering was previously referred to as resonance anomalous x-ray reflectivity (Walker, 1991). However, we will refer it to resonance anomalous surface scattering (RASXS) since it is not only applicable to the reflectivity techniques but also to off-specular CTR measurements and glancing angle in-pane diffraction.

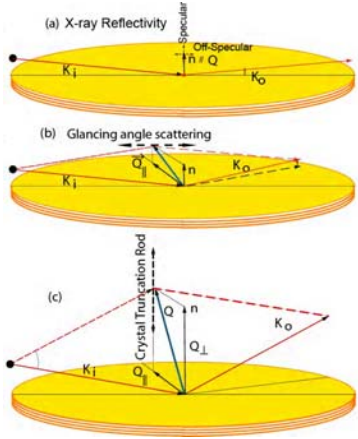

Fig. 1. Experimental Geometries of Surface x-ray scattering.

In this paper, following the experimental description in Sec. 2, we will give a brief review of a RASXS technique and its application to anodic oxide formed on platinum surfaces. It will then be followed by discussions on new prospects of the technique by incorporating polarization-dependence to this technique. The polarization dependence measurements in principle open up completely new applications of RASXS. We will show that RASXS is applicable to all three types of surface scattering including glancing angle x-ray diffraction, well beyond the reflectivity regime as Walker et al. (1991) originally envisioned.

**2. Experimental**

Experiments were performed at beam lines ID11-D and ID12-B at the Advanced Photon Source (APS), Argonne National Laboratory. The monochromators of the beamlines are cryogenically cooled Si(220) with an energy resolution of ~1.5 eV and Si(111) with an energy resolution of ~2 eV, respectively. Both beam lines are equipped with six-circle diffractometers (You, 1999). The polarizations were defined by constraining the surface normal either to the vertical plane (σ-polarization) or to the horizontal plane (π-polarization) containing the incoming x-rays.

Two types of background subtractions were always employed in order to ensure that only elastically scattered x-rays are detected. First, we used energy-dispersive detectors, such as CdZnTe or Si PIN diode detectors made by AMTEK at Bedford, MA USA or Si avalanche diode Vortex detectors made by Radiant Detector Technologies at Northridge, CA USA. We find that the new Vortex EX detectors exhibit significantly narrower energy resolution and higher counting rate for our applications. For the reason, most recent data are all taken with Vortex EX detectors. The detector signals were fed to an amplifier typically with 0.25 or 0.125 μsec shaping time and base-line restoration, and the amplifier output then subsequently fed to a pulse height analyzer (typically, AIM module by Canberra at Meriden, CT USA). In this way, we were able to discriminate the most unwanted fluorescence background scattering. The shaping time was chosen to compromise the high count rate and suitable energy resolution to discriminate the fluorescence background scattering. Second, we subtracted diffuse background for every data point of the scans by subtracting the duplicated measurements off by ~0.5° from the surface peak as our typical CTR and reflectivity measurements were done. When both types of background subtraction are done, only the true elastic x-rays scattered from the surface or interface are counted. When the surface scattering signal is sufficiently strong compared to the fluorescence background, the first type of energy-dispersive discrimination is not necessary.

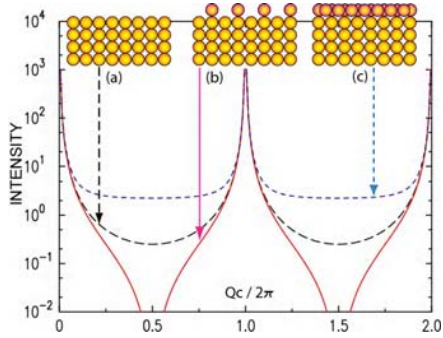

Fig. 2. Calculated surface scattering intensities for three different surface coverage: (a) no surface layer, (b) ½ monolayer, and (c) 2 monolayers.

## 3. Introduction to Resonance Anomalous Surface X-ray Scattering

Ordinary surface x-ray scattering can be modeled by simple kinematic scattering formula. For convenience, we can divide the scattering amplitude to surface component and the bulk component. This division is purely artificial since the surfaces of the real samples are melded parts of the bulk. Nevertheless, one can make this division without loss of any generality as long as the surface part encompasses all the details of the deviation from the bulk. Let us denote it $S(\mathbf{Q},E)$. Then the amplitude from the bulk component can simply be modeled by the CTR amplitude of a half infinite crystal with an ideal surface. Let us denote it $B(\mathbf{Q},E)$. This is generally a complex number which has strong amplitudes only at the bulk reciprocal lattices. They are streaks or rods of amplitude whose directions are normal to the surface and whose magnitudes are rapidly decreasing from the reciprocal lattices. The total amplitude, $A(\mathbf{Q},E)$, then becomes:

$$A(\mathbf{Q},E) = S(\mathbf{Q},E) + B(\mathbf{Q},E) \qquad (1)$$

In this way, we can discuss the bulk term and surface term separately in the following subsections. We can also easily examine the surface sensitivity of the technique by demonstrating the interference between the bulk and surface terms of simple models.

### 3.1. Scattering amplitude from the half-infinite crystal

For simplicity, we will restrict ourselves to specular reflections. Then we can reduce the bulk term, $B(\mathbf{Q},E)$, to a half-infinite one-dimensional (1d) crystal as

$$B(\mathbf{Q},E) = f(\mathbf{Q},E) \cdot b(\mathbf{Q}) \qquad (2)$$

where $f(\mathbf{Q},E)$ is the structure or form factor and $b(\mathbf{Q})$ is the amplitude from a half-infinite δ-function array. Note that $b(\mathbf{Q})$ is independent of energy. Within the 1d approximation, we can evaluate $b(\mathbf{Q},E)$ by employing a usual mathematical tick that Q is Q−iε but ε→0. In real scattering experiment, ε manifests as absorption or extinction. Then,

$$b(\mathbf{Q}) = 1 + e^{-i\mathbf{Q}c} + e^{-2i\mathbf{Q}c} + \cdots \qquad (3)$$

$$= \frac{1}{1 - e^{-i\mathbf{Q}c}} \qquad (4)$$

$$= \frac{1}{2} - i \frac{\cos(Qc/2)}{2\sin(Qc/2)} \qquad (5)$$

Note that $\mathbf{Q} = 2\pi/c$ is a Bragg condition and $b(\mathbf{Q})$ diverges. If $\mathbf{Q}$ is π/c, however, $b(\mathbf{Q})$ is ½. We call this case an anti-Bragg condition. At this condition, the scattering amplitudes from each layer is alternatively out-of-phase as one can see in Eq. (3), but the scattering from one half monolayer effectively remains once all the amplitudes from layers are added together. The intensity from such a half-infinite surface is illustrated by the long dash (black) lines in Fig. 2(a).

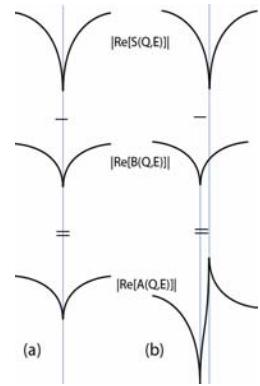

Fig. 3. A schematic representations of $A(\mathbf{Q},E)$, $S(\mathbf{Q},E)$, and $B(\mathbf{Q},E)$ for an ordinary case (a) and for the surface atoms with a larger resonance energy (b).

## 3.2. Simple examples of surface layers

Let us now examine the effect of the surface layers to the total scattering amplitude $A(\mathbf{Q},E)$. In Fig. 2(b), a half-monolayer surface layer is added to the bulk. The scattering intensity is shown with solid (red) lines. Note that the intensity at the anti-Bragg condition is zero because the surface contribution cancels the bulk contribution. In Fig. 2(c), a two-monolayer surface layer is added to the bulk. In this case (short dashed lines), the intensity is $(2-\frac{1}{2})^2$. From this illustration, we can see the sensitivity of the surface scattering intensity, in particular, of the anti-Bragg intensity to the surface layers. A similar exercise can be made for surface layer expansions and demonstrate the sensitivity to the position of the surface layer (Robinson and Tweet, 1992).

The surface scattering amplitude from the surface layers at or near anti-Bragg conditions is in general out-of-phase with the bulk contribution. For this reason, the surface atoms with different chemical states, otherwise identical to those of the bulk, can be determined. The simple diagrams shown in Fig. 3 can be used to illustrate the effect of the difference in chemical states on the anti-Bragg intensity. Fig. 3(a) illustrates an ordinary surface where no chemical shift of the surface atoms is expected. Since the resonance energy is the same for every atom, the resulting amplitude, $A(\mathbf{Q},E)$, shows a simple resonance behavior. However, if the surface atoms have a chemical shift as shown in Fig. 3(b), $A(\mathbf{Q},E)$ will show a significantly modified shape of resonance.

## 3.3. Experimental example

Platinum is one of the most important metals for its high catalytic activities for energy conversion process and environmental control of automobile or industrial exhausts. Its properties are in many ways unique, but its anodic oxidation behavior is particularly so and has been studied for many years. By combining CTR measurements (You, *et al.*, 1994) and RASXS (Chu, *et al.,* 1999), we determined the structure and chemical states of Pt(111) surface platinum in various oxidation states. Briefly, we determined that there are three distinctly different oxidation states shown in Fig. 4 in function of electrode potential. (a) Up to ~0.9 V vs. reversible hydrogen electrode (RHE) there was little change in the anti-Bragg $(0\ 0\ 1.5)_{hex}$ x-ray scattering intensity in spite of the obvious electrochemical signatures (not shown) at around 0.8 V. This means that chemisorption of oxygen or OH occurs but has no effect on the surface Pt atom positions as shown in Fig. 4(a). (b) Between ~0.9 and ~1.2 V (RHE), the anti-Bragg intensity was found to decrease sharply with the accompanying electrochemical current spike. We determined that as much as one third monolayer of the top Pt atoms exchanged places with the chemisorbed oxygen. This place exchange was completely reversible upon reduction. Full CTR analysis indicated the rearrangement of Pt atoms by the place exchange as schematically shown in Fig. 4(b). (c) Oxidation at >1.2 V (RHE) shows no change of Pt surface structure from the previous case as measured by CTR. However, either the reduction of this oxide or simply holding the potential for several minutes irreversibly roughens the surface Pt structure.

In case (a) where oxyspecies chemisorbed on the surface, the RASXS measurements found no measurable chemical shift despite the expected chemisorption. We concluded that in this case the chemical shift is too small to be measured with this technique. In case (b), we found that the measured RASXS spectra were easily modeled by the oxide monolayer (1/3 Pt layer and 2/3 oxygen). The fit value for the resonance energy shift of the top 1/3 monolayer Pt atoms was the surprisingly large value of 9(2) eV. Similarly, we were also able to model the case (c) with the resonance energy shift of the top *two* layers (1/3 first and 2/3 second layers). We concluded that the top two layers of Pt and oxygen atoms form a bilayer oxide of $PtO_2$ when the electrochemical potential exceeds ~1.2 V.

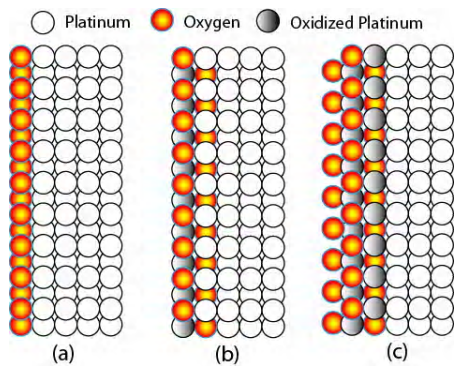

Fig. 4. Three models of Pt surface oxides are shown: (a) Oxygen chemisorbed, (b) Top Pt atoms are oxidized, and (c) Top two Pt layers are oxidized.

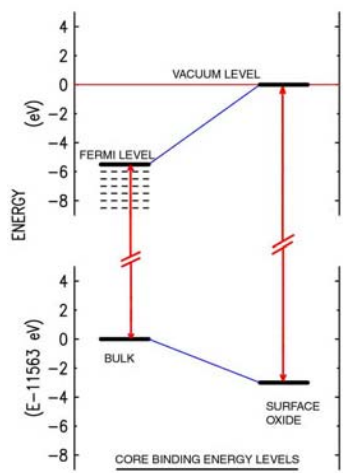

Fig. 5. A schematic diagram illustrating the energy levels of ground states and intermediate states.

The surprisingly large shift in the resonance energy seemed unreasonable at first. However, our modeling of the data yielded unambiguously this seemingly unreasonably large shift. Noting that the resonance energy is the energy *difference* between the ground and the lowest unoccupied intermediate states (Chu, 1999), we propose the following explanation for the large resonance energy shift (See Fig. 5). In case of bulk platinum in metallic state, the lowest unoccupied intermediate state is at the Fermi level. In the case of oxide layer(s), the Fermi level may fall at the insulating band gap, thus the lowest intermediate state available is at the vacuum level. Since the work function of platinum metal is ~6 eV, the actual binding energy shift of the oxidized Pt atoms is ~3 eV. This value is not unreasonable for the $L_{III}$ core binding energy shift upon oxidation.

We have discussed this example to demonstrate that RASXS can be used to determine the chemical state of the surface atoms relative to those of the bulk. In this example, all the measurements were performed in σ-polarization, i.e., the surface was parallel to the polarization vector of the linearly polarized synchrotron x-rays. The σ-polarization was chosen merely for an experimental convenience without recognizing the potential importance of the polarization dependence at the time of these early experiments.

**4. Polarization-dependence**

Templeton and Templeton (1980) have reported the observation of x-ray dichroism. They performed x-ray absorption experiments with crystalline vanadyl bisacetylacetonate through the vanadium K edge with linearly polarized synchrotron radiation. They have shown that the interaction of x-rays with non-cubic symmetries can lead to polarization-dependent near-edge absorption. Their study was motivated by the polarization-dependent extended x-ray absorption fine structure (EXAFS) observed on Se K edge in $WSe_2$ (Heald and Stern, 1977) and the study of bromine on graphite (Heald and Stern, 1978). However, it was the first study which deliberately demonstrated x-ray dichroism in an absorption experiment (*f"*) and also it was the first experiment on diffraction dichroism in scattering experiment (*f'*) based on the Kramers-Kronig relation. Subsequently, they (Templeton and Templeton, 1985) demonstrated that the dichroism can indeed be extended to polarization-dependent diffraction from single crystals with non-cubic symmetries. The measurement was done on $K_2PtCl_4$ single crystals of highly anisotropic layered symmetries by tuning x-ray through the Pt $L_{III}$ edge.

An atom on a surface or interface experience a similar anisotropic environemnt as in highly anisotropic 3d crystals such as $K_2PtCl_4$. For this reason, one can suppose that a similar diffraction dichroism from the surfaces and interfaces. In this section, we will show that one can indeed apply the Templeton scattering (Templeton and Templeton, 1985) to surface and interfaces. That is, we will show that RASXS can have strong polarization dependence. The polarization dependence will be demonstrated for several cases of Pt(111) surfaces by measuring through the Pt $L_{III}$ edge.

*4.1. Resonance Anomalous scattering cross section*

At or near resonance condition for a specific absorption edge of an atom, scattering amplitudes from an atom can be divided into a resonance term of the edge and non-resonance terms. Then the resonance term is expressed in a usual scattering equation of real and imaginary parts multiplied by Thompson scattering cross section and absolute square of transition matrix elements. The matrix elements are written as $\langle A | e^{i\mathbf{k}\cdot\mathbf{r}} \mathbf{p} \cdot \mathbf{\varepsilon} | I \rangle$ where *A* and *I* stand for ground and intermediate states, respectively. Evaluation of this matrix typically involves numerical computation. One of the most successful numerical approaches is the *ab initio* multiple-scattering approximation (Ankudinov and Rehr, 2001). We will use this approach to compare the computational results to our experimental results. However, we will also have elementary theoretical discussion in the last section of this paper to illustrate the first-principle explanation of the polarization dependence seen in our experiments.

Since $\mathbf{k}\cdot\mathbf{r}$ is not small in a typical RASXS experiment using hard x-rays, typical selection rules will not apply. This adds some complication to the problem and actual computation of the transition matrix elements must be performed in more general forms. The calculations (Ankudinov and Rehr, 2001) are limited to the forward scattering, the error for the momentum transfer less than 2 $Å^{-1}$ used in our experiments is expected small in modeling of ordinary anomalous scattering experiments or RASXS experiments. However, it is desirable to develop such computational schemes for non-zero momentum transfer for accurate simulation. For this reason, the discussions of our data given in this paper will remain qualitative at this time although we demonstrate the potential usefulness of the technique.

*4.2. Platinum surface oxide*

As reviewed in Sec. 3.3, Chu, et al. (1999) have shown that the chemical environment of the platinum surface oxidation can be studied with RASXS where the measurements were performed with σ-polarized synchrotron x-rays. Here, as the first example, we will compare and show that π-polarization measurements yield very different scattering spectrum from that of σ-polarization. The measurements are made for the oxide schematically shown in Fig. 4(b). We present the raw data without geometric and background corrections in Fig. 6. Even without any form of data reductions or corrections, it is clear that they are qualitatively and significantly different. The most notable difference is the sharply rising peak right above the edge which we identify as the result of d-band bonding to oxygen atoms. Since the Pt-O bond direction is along the surface normal in this example, we attribute the strongly enhanced peak several eV above the edge in the π-polarization to the overlap of Pt-O bonds and polarization vector in the transition matrix elements. We will revisit this polarization effect in the simpler case of Pt(111)/CO in later sections. Here we merely demonstrate that a large difference is real and can be seen in the system which we are familiar with.

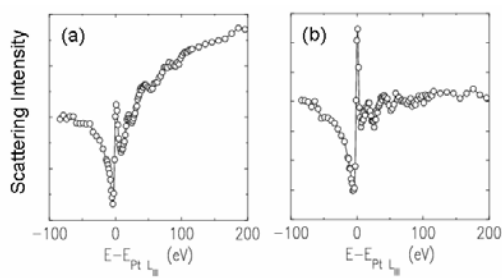

Fig. 6. Pt L$_3$ σ- (a) and π- (b) polarization RASXS data measured at the anti-Bragg position for Pt(111) surface oxide.

### 4.3. Sulfate adsorbed on Pt(111) surface

Surface electrochemistry studies are generally performed on the single crystal surface immersed in an electrolyte. One of the most widely used electrolytes is sulfuric acid. A typical current-voltage curve is shown in Fig. 7. A main feature of the current-voltage curve in the sulfuric acid is the adsorption of (bi-) sulfate to the surface. The adsorption of the (bi-) sulfate shows an extremely sharp electrochemical signature at 0.22 V (vs. Ag/AgCl, add 0.25 V to make comparable to the voltage unit vs. RHE) in the current-voltage curve. The sharp features are known to occur due to the ordering (the positive peak) and disordering (the negative peak) of (bi-) sulfates. It is reported that they are ordered in the potential range between 0.22 and 0.45 V and replaced by hydrogen adsorption below 100 mV (Funtikov et al., 1997).

We chose two potentials, 50 mV and 400 mV, for our measurements. At 50 mV, the (bi-) sulfate is not expected to be on the surface (instead, hydrogen is expected on the surface at this potential), and at 400 mV, it is expected to be on the surface and ordered. However, unlike the anodic oxide that we have discussed earlier, the (bi-) sulfates do not adsorb strongly and are not expected to form chemical bonds to surfaced Pt atoms. The measured RASXS data are shown in Fig. 8. The circles and squares represent the scans made at 400 mV with σ-polarization and π-polarization, respectively. The close-up view of the difference is shown in the inset. The π-polarization scan shows clearly the features that are not present in the σ-polarization scan. The triangles represent the π-polarization scan at 50 mV. This scan does not show the feature shown for π-polarization 400 mV scan and is close to the scans for hydrogen adsorbed surfaces measured in non-absorbing HClO$_4$ electrolyte (Chu, 1999). These scans indicate that the feature seen in the π-polarization at 400 mV scan comes from the adsorbed (bi-) sulfate. We believe that the difference seen here can be used to study the structure and chemistry of absorbates in electrochemical interfaces even if the species are not chemically bonded to the substrates. If theories such as FEFF8.2 (Ankudinov, 2001) are further developed, the polarization dependence seen in our data can be quantitatively analyzed to yield structure and chemical information otherwise difficult to obtain.

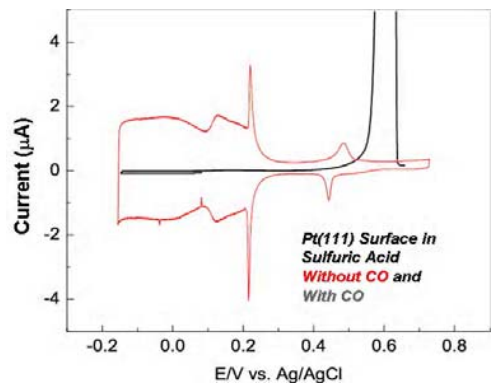

Fig. 7. Typical current-voltage curve of Pt(111) in sulfuric acid showing several electrochemical signatures. The featureless curve is observed when the surface is poisoned by CO adsorption.

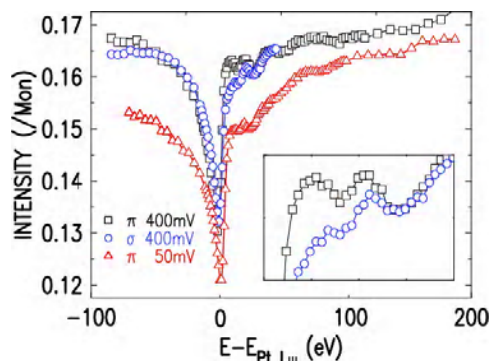

Fig. 8. Scans made for sulfate adsorbed on Pt(111) at Pt L$_{III}$ edge. The inset shows a close up view of the difference in the two polarization scans made at 400 mV. The 50 mV scan is vertically displaced for clarity.

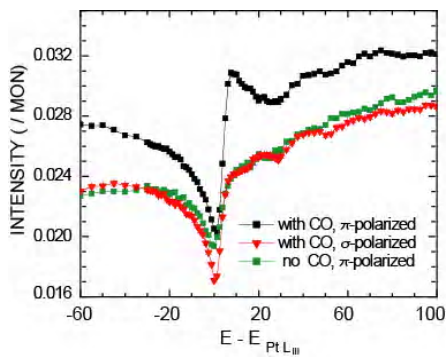

Fig. 9. Three energy scans through Pt $L_{III}$ edge at the anti-Bragg condition. The p-polarization scan made for CO/Pt(111) shows a strong enhancement several eV above the resonance energy.

### 4.4. Carbon monoxide adsorbed on Pt(111)

Monolayers of CO adsorbed on Pt(111) surface have been a model system for generations of surface scientists. There have been numerous studies and reviews. They are too many to list here, and a review on the subject is beyond the scope of this paper. Recently, interest on CO monolayers on platinum surfaces has been reinvigorated by fervid research activities for mitigation of CO poisoning problems in fuel cell catalysts. Most CO interactions with catalysts occur near atmospheric or ambient pressure and/or under electrolytic conditions. In these conditions, most surface science techniques developed and used in high or ultra-high vacuum conditions cannot be used because of the relatively short electron mean free paths in ambient gas or electrolytic conditions. Therefore, RASXS for buried interfaces under high-pressure gases, liquids, or other solids can be an extremely valuable tool.

We chose CO/Pt(111) because it is well known and standard system. It is a good system to develop a new technique. For this reason, we will discuss this system in more detail than the two previous examples. The effect of CO on electrochemical activity is well demonstrated in Fig. 7. When CO is absorbed, all the features in the current-voltage curve essentially disappear as shown by the black solid line except for the sharp rise at 600 mV due to oxidation of the absorbed CO. Within the potential range shown in Fig. 7, two phases of CO layers are known to exist; (2×2)-3CO (Markovic, *et al.*, 1999) below 350 mV and (√19×√19)R23.4°-13CO (Tolmachev, *et al.*, 2004) between 350 mV and 600 mV. Here we will mainly discuss the (2×2) structure for our example.

In Fig. 9, we show three energy scans at the anti-Bragg position (0 0 1.5) though the Pt $L_{III}$ made at −200 mV; a π-polarization scan with the (2×2) CO present on the surface (black squares); a σ-polarization scan (triangles) and a π-polarization scan *without* CO on the surface (green squares). The scans are shown again without geometric corrections or reductions and should be compared qualitatively. The main differences between the two polarizations have been reproduced many times in different CO-saturated electrolytes and also in CO gas. The raw data are presented here to show that the difference is significant and can be seen without any type of data manipulation.

The lower two scans, the σ-polarization with CO (green squares) and the π-polarization without CO (triangles), are quite similar each other but qualitatively different from the π-polarization scan *with* CO (black squares). A difference can also be noted between the lower two scans. Without CO on the surface, the resonance dip does not go as deep as the one with CO. The most pronounced difference is between the top scan (black squares) and the lower two scans. The sharply rising feature, similar to the so-called white line in a near-edge absorption spectrum, several eV above the resonance energy is unique for the π-polarization scan *with* CO absorbed on the surface. Like the two previous examples, the surface platinum oxide and sulfate on platinum, we speculate that this is also due to the bond directions of Pt-C (inter-bond) and C-O (intra-bond). This is quite analogous to the Templeton scattering since the excited Pt core electron will experience a strong anisotropic environment due to the empty states present in intra-bonds of CO molecules and inter-bonds between CO and the platinum atom(s) that it is bonding.

More dramatic differences can be seen when the polarization scans are made at the respective superlattice reflections (not anti-Bragg) of the (2×2) and (√19×√19) structures. The scans are shown in Fig. 10. The strong superlattice reflection of the (2×2) structure occurs at (0.5 0.5 0) and that of the (√19×√19) structure appears at (3/19 14/19 0). The actual measurements are done at glancing angle geometry with L=0.2. Again, no corrections or reductions are made to the data shown in Fig. 10. One can see the fine structures in the scans. Moreover, differences between polarization scans in both structures are striking. We believe that the difference in the fine structures will enable us to study the two-dimensional short-range, local, and chemical environments when we have theoretical or computational tools to simulate these types of data.

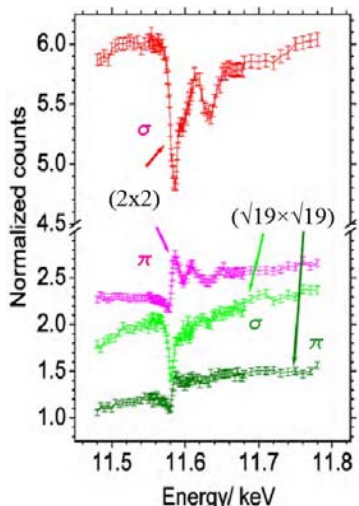

Fig. 10. The energy scans through Pt $L_{III}$ edge at superlattice peaks. σ- and π- polarization scans were made for (2×2) and (√19×√19) structures.

## 4.5. Theoretical considerations for the dichroism in RASXS of CO/Pt(111)

In this section, we will consider two theoretical approaches in attempt to understand the scans shown in Fig. 9. One method is the first-principle approach where one computes the density of states using density functional theory and then computes the transition matrix elements to the calculated density of states. The other approach is the multiple-scattering methods where the core electrons are ejected by the x-ray photon and undergo multiple scattering through the nearby atoms before being reabsorbed. Although the first-principle approach is more elegant, the multiple scattering approaches are currently more practical, more importantly, significantly developed over the last decade (Ankudinov and Rehr, 2001). For this reason, we will discuss the latter approach first and compare to our data. Later we will have more discussion on the first principle approach.

### 4.5.1. Multiple scattering approach

We have employed the FEFF8.2 code, the latest version of the series (Ankudinov and Rehr, 2001), in calculating the scattering intensity. In order to illustrate the approach we used the smallest unit cell containing the essential atoms in the system. The result is shown in Fig. 11. The calculations with larger cells produce essentially similar results (not shown) with some differences only in details. The unit cell of four surface Pt atoms and three adsorbed CO molecules used in the computation is shown as an inset to Fig. 11(a). The open circles are Pt atoms, the blue (directly above Pt) and red circles are carbon and oxygen, respectively. Note that one CO molecule is linearly bonded to a Pt atom on atop site and two others are bonded to three Pt atoms by occupying the hollow sites. The $L_{III}$ electrons of the Pt atom with the atop CO are excited in the computation. Since the hollow site CO molecules are relatively far away from the Pt atom, it was not surprising to find that the results are largely unaffected by those two CO molecules.

Unfortunately, at this current version, the computation is limited to forward scattering cross sections although the improvement for non-zero momentum transfers is in progress. Nevertheless, the results are sufficiently illuminating and reproduce the most essential features of the data presented in Fig. 9. First, the π-polarization computation with CO on the surface clearly shows an enhanced scattering several eV above the resonance energy. It also shows the deepest resonance in agreement with the data. Comparing the σ-polarization computation with CO and π-polarization (σ-polarization is similar) without CO, we note that the resonance dip is shallowest for the Pt-only case which is also in complete agreement with the data. The computed differences are shown in Fig. 11(b). The differences demonstrate the sensitivity of RASXS to the mono- or submono- layer level changes on the buried interfaces and also the diffraction dichroism which is potentially sensitive to the bonding and chemical configurations of the absorbates.

While the essential features are clearly shown, the realistic computation of the scattering intensity is far from simple. Even if the computation is extended to an arbitrary momentum transfer, a realistic choice of a cluster with sufficient size and most sensible cluster shape must be made. Then the computations of the scattering amplitude for the various atoms in the cluster with different symmetries must be separately performed. The diffraction intensity should be then computed for every x-ray-energy point where measured data are available and recombined in order to obtain the intensity scan comparable to the experimental data. This requires a considerable amount of computing and fit of the computation to the data will not be realistic at least at this point. Therefore, we remain at qualitative comparison as we have done here.

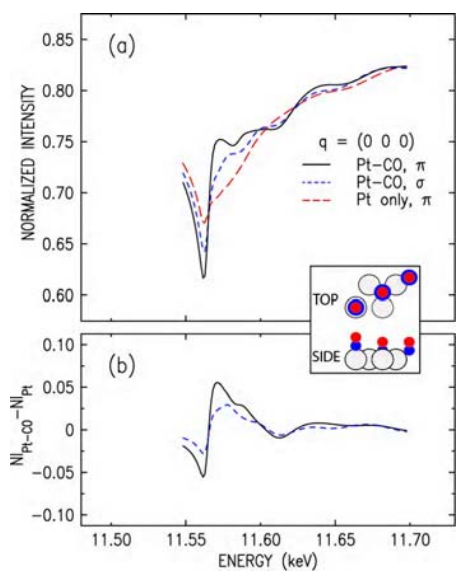

Fig. 11. Computation of the resonance scattering for σ- and π-polarizations (a) and their differences to the bare Pt(111) surface is shown in (b) for discussion. The cluster used in the computation is shown as the inset.

*4.5.2. Origin of polarization dependence: Elementary view*

Resonance scattering from a collection of atoms can be expressed in terms of transition matrix elements and the resonance energy and incoming x-ray energy. A complete first-principle evaluation of this equation is obviously very complex and we will not be able to cover in this paper. However, it is valuable to perform elementary expansion of the matrix element to see the origin of the polarization dependence. For this, we would like to separate the CO specific transition matrix elements from all other resonance term. That is, the double summation over the ground state and intermediate states will be separated to two terms; one that is reduced to the Pt $L_{III}$ edge and to the states associated with intra- and inter-bonding of CO, respectively and the other for the remaining other resonance terms. We assume that CO absorption does not significantly affect Pt atoms except for the one directly bonded to CO. Bonding configurations are known (Wong and Hoffmann, 1991) and the lowest unoccupied state is $2\pi^*$ state. The bonding and anti-bonding of this state and other higher energy states will be the basis for the CO induced states. Then the cross section can be written as

$$f(\mathbf{Q},\omega) = \sum_i e^{i\mathbf{Q}\cdot\mathbf{r}_i} \sum_{I_{CO}} \frac{r_0 \left|\left(e^{i\mathbf{k}\cdot\mathbf{r}}\mathbf{p}\cdot\mathbf{\varepsilon}\right)_{I_{CO}L_{III}}\right|^2}{m\hbar}$$
$$\times\left(-\frac{2}{\omega} + \frac{1}{\omega_{I_{CO}L_{III}} - \omega - i\frac{\Gamma}{2\hbar}} + \frac{1}{\omega_{I_{CO}L_{III}} + \omega}\right)$$
$$+ \sum_i e^{i\mathbf{Q}\cdot\mathbf{r}_i} \sum_A \sum_I \frac{r_0 \left|\left(e^{i\mathbf{k}\cdot\mathbf{r}_i}\mathbf{p}\cdot\mathbf{\varepsilon}\right)_{IA}\right|^2}{m\hbar}$$
$$\times\left(-\frac{2}{\omega} + \frac{1}{\omega_{IA} - \omega - i\frac{\Gamma}{2\hbar}} + \frac{1}{\omega_{IA} + \omega}\right)$$

where $\mathbf{k}$ should be noted, to be accurate, as $\mathbf{k}_{in}$ and $\mathbf{k}_{out}$. For simplicity of the equations, we consider only the forward scattering cross sections and ignore the difference.

In order to further simplify the case, we will consider the resonance anomalous scattering from a single Pt-CO bonded molecule with $\sigma$- and $\pi$- polarizations. In this case, the matrix elements associated with the $L_{III}$ edge can be written as at the top of the next column where the first equality can be obtained from the identity, $\vec{p} = \frac{i}{\hbar}[H,\vec{r}]$ and by assuming that the wave functions can be rewritable in a cylindrical coordinate with the z axis along the bond direction. Also, note that the polarization vector $\mathbf{\varepsilon}$ is expanded to the component along z direction ($\varepsilon_z$) and normal to z direction ($\varepsilon_\rho$). Now then, in the second equality, we can divide two terms each of which depends only on

$$\left(e^{i\mathbf{k}\cdot\mathbf{r}}\mathbf{p}\cdot\mathbf{\varepsilon}\right)_{I_{CO}L_{III}}$$
$$= im\omega \int d\phi dz d\rho \rho \Phi_{L_{III}} Z_{L_{III}} \Gamma_{L_{III}}$$
$$\times \left[e^{ik_\rho \rho \cos\phi} e^{ik_z z}(\varepsilon_\rho \rho \cos(\phi - \pi/2) + z\varepsilon_z)\right]$$
$$\times \Phi_{CO} Z_{CO}(z - z_{CO})\Gamma_{CO}$$
$$= im\omega\varepsilon_\rho \int d\phi \Phi_{L_{III}} \Phi_{CO}$$
$$\times \int d\rho \rho \Gamma_{L_{III}} \Gamma_{CO}\left[e^{ik_\rho \rho \cos\phi}\rho \cos(\phi - \pi/2)\right]$$
$$\times \int dz Z_{L_{III}}\left[e^{ik_z z}\right]Z_{CO}(z - z_{CO})$$
$$+ im\omega\varepsilon_z \int d\phi \Phi_{L_{III}} \Phi_{CO}$$
$$\times \int d\rho \rho \Gamma_{L_{III}} \Gamma_{CO}\left[e^{ik_\rho \rho \cos\phi}\right]$$
$$\times \int dz Z_{L_{III}}\left[e^{ik_z z}z\right]Z_{CO}(z - z_{CO})$$

either $\varepsilon_z$ or $\varepsilon_\rho$. Therefore, only one of the two terms remains depending on the polarization. Now in this equation, one can see the origin of the polarization dependence. In our experiments, the first term is small or zero while the second term has an appreciable value in the integral. The evaluation of the overlap integrals can in principle be numerically performed. However, note that the transition matrix will not follow the usual selection rules of dipole transition because the $\mathbf{k}\cdot\mathbf{r}$ is not small compared to $2\pi$ and the exponential term $e^{i\mathbf{k}\cdot\mathbf{r}}$ cannot be approximated by 1. Development of numerical evaluation of such integrals will be a challenging but important task. We believe that the initial step can be taken using density function theory.

## 5. Conclusion

We have shown that RASXS and its polarization dependence has the potential to be used to study local and chemical information of the buried interfaces, otherwise unobtainable by any other means. We have shown that surface diffraction and RASXS combined can determine the structure and chemical states of the buried interfaces. We also have shown that the $\pi$-polarization RASXS is very sensitive to surface absorbates even when they are not chemically bonded. We attribute the sensitivity to the anisotropic local environment of Pt atom of the surface analogous to that of Pt atoms in $K_2PtCl_4$ with highly anisotropic symmetry used in Templeton scattering (Templeton and Templeton, 1985).

**Acknowledgement**

We would like to thank Drs Guy Jennings and Klaus Attenkofer for useful discussion and assistance during the experiments. This work and the use of the Advanced Photon Source was supported by the Office of Basic Energy Sciences, U.S. Department of Energy under contract no. W-31-109-ENG-38.